\documentclass[11pt,a4paper]{article}

\usepackage[T1]{fontenc}
\usepackage[utf8]{inputenc}
\usepackage[english]{babel}
\usepackage{csquotes}

\usepackage[
  a4paper,
  margin=1in,
  includeheadfoot,
  top=0.90in,
  headsep=20pt
]{geometry}

\usepackage{microtype}
\usepackage{parskip}
\usepackage{setspace}
\usepackage{ragged2e}
\AtBeginDocument{\justifying}

\usepackage{newtxtext}
\usepackage{newtxmath}

\usepackage{amsmath}
\usepackage{mathtools}

\usepackage[dvipsnames]{xcolor}
\definecolor{Accent}{HTML}{1D4ED8}
\definecolor{Accent2}{HTML}{0F766E}
\definecolor{Dark}{HTML}{111827}
\definecolor{SoftBG}{HTML}{F3F6FB}
\definecolor{Rule}{HTML}{CBD5E1}
\definecolor{Muted}{HTML}{475569}

\usepackage{hyperref}
\hypersetup{
  colorlinks=true,
  linkcolor=Accent,
  citecolor=Accent2,
  urlcolor=Accent
}
\usepackage[nameinlink,capitalise,noabbrev]{cleveref}

\usepackage{graphicx}
\usepackage{float}
\usepackage{caption}
\usepackage{booktabs}
\usepackage{tabularx}
\usepackage{array}
\newcolumntype{Y}{>{\raggedright\arraybackslash}X}
\usepackage{siunitx}
\usepackage{enumitem}
\setlist[itemize]{topsep=4pt,itemsep=2pt,leftmargin=*}
\setlist[enumerate]{topsep=4pt,itemsep=2pt,leftmargin=*}

\usepackage[most]{tcolorbox}
\tcbset{
  enhanced,
  colback=SoftBG,
  colframe=Rule,
  boxrule=0.8pt,
  arc=2mm,
  left=10pt,right=10pt,top=8pt,bottom=8pt
}
\newtcolorbox{AbstractBox}{
  colframe=Rule,
  colback=SoftBG,
  title=\textbf{Abstract},
  fonttitle=\color{Dark},
  coltitle=Dark
}

\usepackage{titlesec}
\titleformat{\section}{\Large\bfseries\color{Dark}}{\thesection}{0.6em}{}
\titleformat{\subsection}{\large\bfseries\color{Dark}}{\thesubsection}{0.6em}{}
\titlespacing*{\section}{0pt}{0.6em}{0.35em}
\titlespacing*{\subsection}{0pt}{0.5em}{0.25em}

\usepackage{fancyhdr}
\renewcommand{\headrulewidth}{0.5pt}
\renewcommand{\headrule}{\hbox to\headwidth{\color{Rule}\leaders\hrule height \headrulewidth\hfill}}
\usepackage[
  backend=biber,
  style=numeric-comp,
  sorting=none,
  maxbibnames=99,
  doi=true,
  url=true
]{biblatex}
\usepackage{fontawesome5}
\newcommand{\orcidicon}{\faOrcid}

\newcommand{\authorname}[4]{%
  \textbf{\color{Dark}#1}\textsuperscript{#3}%
  \href{https://orcid.org/#2}{\textcolor{Accent}{\,\orcidicon}}%
  \ifnum#4=1\relax\textsuperscript{*}\fi
}

\newcommand{\corremail}[1]{%
  {\small\color{Muted}\textsuperscript{*} Corresponding author: \href{mailto:#1}{#1}\par}
}

\newcommand{\papertitle}{Concentration and Distribution of Container Flows in Mauritania’s Maritime System (2019–2022)}
\newcommand{\paperkeywords}{maritime container transport, route concentration, port dependency, trade logistics vulnerability, structural asymmetry, West Africa}

\makeatletter
\renewcommand{\maketitle}{%
  \begingroup
  \thispagestyle{empty}

  \vspace*{-1.20cm}

  {\color{Accent}\rule{\linewidth}{1.4pt}}\par
  \vspace{0.15cm}

  {\bfseries\LARGE\color{Dark}\papertitle\par}
  \vspace{0.22cm}

  {\color{Rule}\rule{\linewidth}{0.5pt}}\par
  \vspace{0.22cm}

  \begin{center}
    {\normalsize
      \authorname{Mohamed Bouka}{0009-0005-8502-9885}{1}{1},
      \authorname{Moulaye Abdel Kader Ould Moulaye Ismail}{0009-0005-6738-9975}{1}{0}
      \par
    }

    \vspace{0.12cm}

    {\small\color{Muted}
      \textsuperscript{1}Research Unit of Governance of Institutions, Faculty of Economics and Management, University of Nouakchott, Nouakchott, 5026, Mauritania \\
      \par
    }

    \vspace{0.08cm}
    \corremail{mohamedbouka50@gmail.com}
  \end{center}

  \vspace{0.15cm}
  \endgroup
  \pagestyle{fancy}
}
\makeatother

\usepackage{lipsum}

\begin{document}
\maketitle

\begin{AbstractBox}
Small, trade-dependent economies often exhibit limited maritime connectivity, yet empirical evidence on the structural configuration of their container systems remains limited. This study analyzes route concentration and node distributions in Mauritania’s maritime container system during 2019--2022 using shipment-level data measured in forty-foot equivalent units (FFE). Routes, origin nodes, destination nodes, and industries are represented as FFE-weighted probability distributions, and concentration and divergence metrics are used to assess structural properties. The results show strong corridor concentration across the seven observed routes ($\mathrm{HHI}=0.296$), with the top three accounting for approximately 84\% of total FFE. Node structures differ by direction: imports are associated with a highly concentrated set of destination nodes ($\mathrm{HHI}=0.848$), while exports originate from only two origin nodes ($\mathrm{HHI}=0.567$) and are distributed across a large number of destinations ($\mathrm{HHI}=0.053$). Industry distributions are more concentrated for exports ($\mathrm{HHI}=0.352$) than for imports ($\mathrm{HHI}=0.096$), with frozen fish and seafood accounting for more than 53\% of export volume. Temporal analysis shows that route concentration remains stable over time ($\mathrm{HHI}\approx0.293$--$0.303$), while node distributions exhibit measurable variation, particularly for export destinations (JSD $\approx 0.395$) and import origins (JSD $\approx 0.250$).
\end{AbstractBox}

\vspace{0.15cm}


\section{Introduction}\label{sec:introduction}

Maritime transport underpins contemporary international trade, with containerization forming its dominant logistics architecture. Standardized containers, scheduled liner services, and port network integration have reshaped the spatial organization of trade by linking national economies to global shipping systems through structured service corridors. Within this system, the distribution of container flows across routes and ports is not neutral; it reflects the configuration of maritime service networks, cargo volumes, industrial composition, and logistical connectivity constraints \cite{guo2024analysis,tsantis2026trade,durmaz2024determining}.

For small, trade-dependent economies, the structural organization of container flows is particularly significant. Limited cargo volumes may restrict the availability of direct liner services, increase reliance on transshipment hubs, and concentrate flows along a narrow set of maritime corridors. Such structural characteristics influence routing flexibility, exposure to service disruptions, and the degree of logistical diversification available to domestic industries. Understanding how containerized trade is distributed across routes, ports, and sectors is therefore essential for assessing the configuration of maritime logistics systems in low-volume trade environments \cite{united2024review,xu2024assessing,verschuur2023systemic}.

The practical relevance of such structural dependencies has been underscored by recent geopolitical disruptions affecting the Red Sea corridor. Since late 2023, security threats have led some carriers operating Asia--West Africa services to reroute via the Cape of Good Hope, with implications for transit times and freight costs along corridors used by low-volume trading economies \cite{united2024review}. This context further motivates the empirical characterization of corridor concentration in constrained maritime systems.

While the literature on maritime logistics has extensively examined hub concentration, port competition, network centrality, and operational efficiency in major trading economies, comparatively less attention has been devoted to the structural configuration of container flows in smaller developing contexts. Empirical evidence remains limited regarding how route concentration, port reliance, and sectoral specialization interact within container systems characterized by relatively modest volumes and constrained service portfolios. This gap is particularly relevant for coastal economies in West Africa, where maritime gateways serve as primary trade interfaces yet systematic quantitative analyses remain scarce.

Mauritania offers a pertinent case for such an investigation. The organization of its international merchandise trade is closely linked to its maritime gateways, particularly the Port of Nouakchott, which functions as a central interface for the country’s imported goods and its integration into global shipping networks. Containerized flows connect foreign ports of loading with domestic discharge points through scheduled liner services, as reflected in maritime connectivity and port throughput indicators. These flows reflect the interaction between external liner service structures, the composition of export activities, and domestic import demand patterns. Examining whether Mauritania’s container network is diversified or concentrated, therefore, provides insight into the structural characteristics of its maritime logistics configuration while remaining within the scope of observable transport system data \cite{meridiam_nouakchott_port_2025,unctadstat_maritime_profile_mauritania}.

This study conducts a structural analysis of maritime container flows in Mauritania over the period 2019--2022 using detailed movement data measured in FFE. The analysis focuses on three interrelated structural dimensions: route concentration, port dependency, and industry specialization. In addition, it evaluates differences between import and export configurations and examines temporal stability in network structure across the study period.

The main objective of this research is to analyze the structural configuration of maritime container flows in Mauritania during 2019--2022. This objective is operationalized through the following research questions:

\begin{enumerate}
\item \textbf{Route Concentration:} What is the degree of maritime route concentration in Mauritania's containerized trade?\label{rq:route_concentration}

\item \textbf{Port Dependency:} To what extent do container flows rely on a limited set of foreign ports of loading?\label{rq:port_dependency}

\item \textbf{Industry Specialization:} How are containerized flows distributed across industrial sectors, and does the export structure exhibit concentration patterns distinct from imports?\label{rq:industry_specialization}

\item \textbf{Import--Export Structural Asymmetry:} Are there systematic differences in route usage and container structure between imports and exports?\label{rq:import_export_difference}

\item \textbf{Temporal Structural Stability:} Does the configuration of the maritime container network remain stable over time, or does it exhibit measurable structural shifts during 2019--2022?\label{rq:temporal_change}
\end{enumerate}

This study provides empirical evidence on the configuration of container-based maritime logistics in a small-volume trade environment. The findings offer a structured assessment of concentration, dependency, and distributional patterns within Mauritania’s container network, while also illustrating an analytical approach applicable to comparable maritime contexts.
The remainder of the paper is organized as follows. Section~\ref{sec:related} reviews the related work. Section~\ref{sec:methodology} presents the analytical framework and the concentration, divergence, and stability measures used in the study. Section~\ref{eda} provides an exploratory description of the shipment-level dataset and its distributional properties. Section~\ref{sec:discussion} reports the empirical results and discussion. Section~\ref{sec:conclusion} concludes with limitations and directions for further research.

\section{Related Work}\label{sec:related}

Maritime container transport is a central component of contemporary international trade, linking production and consumption systems through structured shipping corridors and port gateways. Over the past two decades, the global container shipping system has evolved into a highly interconnected network characterized by uneven connectivity and concentration patterns, alliance-driven service structures, and asymmetric trade flows. While extensive research has examined the topology, connectivity, and resilience of the global maritime network, comparatively less attention has been devoted to the structural configuration of container flows in smaller, trade-dependent economies.

Existing studies predominantly focus on large-scale global or regional systems, emphasizing network centrality, hub dominance, alliance structures, and the propagation of disruption. These approaches provide valuable insights into the architecture of the global shipping system but often abstract from the distributional structure of flows within individual national systems, particularly those with modest trade volumes and constrained service portfolios. In such contexts, structural concentration and gateway dependency may have disproportionate implications for trade exposure and logistical vulnerability.

The literature relevant to this study can be organized into four interrelated strands: (i) maritime network structure and topology, (ii) concentration and gateway dominance, (iii) connectivity and trade dependency, and (iv) vulnerability and resilience in shipping networks. Together, these strands provide the conceptual foundation for analyzing route concentration and port dependency as measurable structural properties of container systems.

\subsection{Maritime Network Structure and Topology} \label{topology}

The modeling of maritime container systems as complex networks has shifted empirical analysis from descriptive port-system frameworks toward formal topological characterization. Vessel-movement-based reconstructions of the global liner shipping network indicate persistent hierarchical organization and structural stability over time. Ducruet and Notteboom~\cite{ducruet2012worldwide} show that, despite traffic growth and hub repositioning between 1996 and 2006, the relative configuration of dominant ports remains structurally robust when evaluated through degree and betweenness centrality. Structural polarization persists even as individual nodal rankings adjust.

Subsequent research formalizes these properties within network science. Xu et al.~\cite{xu2020modular} construct a Global Liner Shipping Network of 977 ports connected by 16,680 links and report an average shortest path length of 2.671 and a clustering coefficient of 0.713, consistent with small-world topology. Global efficiency reaches 82.7\% of the fully connected benchmark, while wiring cost accounts for only 1.5\%, indicating a configuration that combines high integration with low spatial redundancy. Community detection identifies seven upper-level modules, and a gateway–hub structural core emerges as the principal integrative mechanism across geographically compact clusters.

Directed representations further reveal asymmetric connectivity patterns. Kang et al.~\cite{kang2022global} construct a 2021 directed GLSN including 564 ports and 2,971 directed links. In-degree and out-degree distributions follow power-law behavior with $R^2 > 0.96$, indicating a heavy-tailed hierarchy. Accessibility measures integrating betweenness, transit impedance, and PLSCI values position a limited group of Asian and European hubs at the upper end of the distribution, reflecting concentration in global accessibility.

Weighted longitudinal reconstructions capture structural evolution over time. Jarumaneeroj et al.~\cite{jarumaneeroj2024evolution} analyze quarterly snapshots between 2011 and 2017 using the Container Port Connectivity Index and modularity-based community detection. Connectivity remains strongly skewed toward East Asian ports, while trading communities reorganize in response to macroeconomic adjustments such as the 2016 Panama Canal expansion. Network restructuring occurs through the reallocation of weighted links rather than through major topological fragmentation.

Network responses to exogenous shocks further illustrate hierarchical differentiation. Using AIS-based vessel movements, Guerrero et al.~\cite{guerrero2022container} document a measurable contraction in weighted connectivity between 2019 and 2020, accompanied by regionally differentiated concentration effects. Large hubs and densely interconnected nodes maintain relatively higher resilience than smaller bridge or transshipment ports, indicating that hierarchical position conditions facilitate shock absorption.

Commodity-layer coupling introduces an additional structural dimension. Ducruet~\cite{ducruet2013network} models maritime transport as a multigraph composed of commodity-specific layers and shows that higher-traffic ports tend to exhibit greater commodity diversity. Traffic allocation across layers remains uneven and is associated with the nodal hierarchy, reinforcing the coexistence of specialization and diversification within the broader maritime structure.

\subsection{Concentration, Vulnerability, and Network Disruption} \label{concentration_vulnerability_network_disruption}

Maritime transportation systems are increasingly examined through the lens of systemic vulnerability, where disruption risk arises not only from isolated node failures but also from structural interdependencies embedded in the network. Calatayud et al.~\cite{calatayud2017vulnerability} conceptualize vulnerability as a function of multiplex configuration, modeling more than 80 liner shipping networks and simulating targeted attacks on seven strategic nodes in the Americas. Their findings indicate that the exposure of freight flows varies depending on a country’s structural position within overlapping service layers. The removal of high-centrality nodes results in disproportionate connectivity losses, suggesting that structural dependencies amplify systemic fragility.

Empirical evidence from vessel tracking further refines this understanding. Using AIS data across 141 disruption events affecting 74 ports and 27 natural disasters, Verschuur et al.~\cite{verschuur2020port} report a median disruption duration of six days and a 95th percentile of 22.2 days. All observed events involve simultaneous multi-port impacts rather than isolated closures. Disruption duration scales with hazard intensity: an additional 1.0 m storm surge or 10 m/s wind speed is associated with approximately two additional days of downtime. Short-term substitution between ports is rarely observed, suggesting that logistical concentration constrains adaptive rerouting capacity. Production recapture rather than spatial diversion emerges as the dominant adjustment mechanism.

Comparative shock analysis reveals differentiated resilience patterns. Notteboom et al.~\cite{notteboom2021disruptions} contrast the 2008–2009 financial crisis with COVID-19, showing that while the financial crisis unfolded sequentially across leading indicators, container volumes, and trade flows, COVID-19 triggered synchronized contractions across operational and financial indicators. Recovery dynamics differ, with COVID-19 showing faster short-term rebounds through adaptive strategies such as blank sailings, slow steaming, and alliance coordination. These findings suggest that resilience is partly endogenous to industry structure and strategic coordination, in addition to network topology.

Climate-related vulnerability introduces an additional structural dimension. Poo and Yang~\cite{poo2024optimising} combine centrality analysis with climate disruption indicators to model route optimization under projected port downtime scenarios. Their framework distinguishes between regional vulnerability linked to nodal centrality and local vulnerability linked to exposure to extreme weather. Ports occupying structurally central positions may exert disproportionate systemic effects even when local hazard levels are moderate.

Large-scale geopolitical shocks provide further evidence of structural adaptation. Cong et al.~\cite{cong2024impact} construct global maritime networks from AIS trajectory data for 20-day windows before and after the Russia–Ukraine conflict. Network connectivity increases by 27.2\%, network scale by 36.6\%, density by 32.4\%, and calculated resilience by 18.6\% following the conflict. Despite localized declines in Black Sea activity, the global network exhibits topological expansion and improved robustness under simulated random attacks, while targeted disruptions continue to generate significant performance degradation.

Across these studies, vulnerability is examined through multiple analytical lenses, including multiplex node removal, empirical downtime measurement, system-level shock comparison, climate-induced disruption simulation, and geopolitical perturbation analysis. Maritime networks are generally characterized by hierarchical concentration and heavy-tailed connectivity, and disruption effects are conditioned by nodal centrality and inter-layer interactions. Adaptive responses operate within constraints imposed by existing concentration patterns.

\subsection{Flow Concentration and Distributional Metrics} \label{flow_oncentration}

Beyond topological connectivity and disruption modeling, an important strand of the literature evaluates maritime systems using market concentration and flow-allocation metrics. In liner shipping, industry concentration is commonly assessed using the Herfindahl–Hirschman Index (HHI) and concentration ratios (CR$_k$). Merk and Teodoro~\cite{merk2022alternative} demonstrate that traditional HHI calculations may underestimate effective concentration when carrier consortia are taken into account. For instance, on the Northern Europe–North America corridor, the conventional HHI was around 1500, whereas the modified HHI (MHHI), accounting for interlinked consortia, approaches or exceeds the 2500 threshold typically associated with high concentration.

Haralambides~\cite{haralambides2019gigantism} discusses the progressive consolidation of liner shipping into three global alliances, highlighting increasing capital concentration and coordinated capacity management. The coexistence of large vessel deployment and alliance coordination implies that service supply and corridor dominance are shaped by coordinated capacity allocation rather than atomistic competition.

At the port-system level, concentration dynamics exhibit spatial variability. Notteboom~\cite{notteboom2010concentration} shows that, although the European container port system experienced phases of deconcentration between 1985 and 2008, container handling remained structurally more concentrated than other cargo segments. The coexistence of multi-port gateway regions and persistent cargo concentration suggests that competitive interaction does not necessarily eliminate dominance patterns.

Empirical applications of HHI to regional port systems further illustrate distributional asymmetries. Nguyen et al.~\cite{nguyen2020competition} calculate HHI values for Southeast Asian container ports and report a score of approximately 0.57 in 2017, indicating high concentration despite emerging competitive dynamics. Their results show that market-share redistribution does not necessarily lead to lower concentration levels.

In non-linear contexts, Lee et al.~\cite{lee2014changing} employ HHI, location quotients, and shift-share analysis to identify deconcentration trends in Korean bulk ports driven by cargo reallocation. Concentration indices are therefore capable of capturing both dominance and structural rebalancing processes.

Connectivity metrics also capture concentration effects indirectly. Fugazza and Hoffmann~\cite{fugazza2017liner} show that the absence of a direct maritime connection is associated with export reductions between 42\% and 55\%, and that each additional transshipment is linked to roughly a 40\% decline in bilateral export value. These findings indicate that the distribution of service connections across corridors materially affects trade performance.

Across these studies, HHI, concentration ratios, and related distributional indicators are used to quantify dominance, competitive structure, and allocation asymmetry. Unlike purely topological measures, these indices explicitly model the distribution of capacity or traffic shares across nodes or routes.

The literature on maritime systems therefore spans three analytical domains: topological structure, disruption resilience, and market concentration. While network science captures integration and modularity, and resilience studies model shock propagation, concentration metrics quantify how traffic or capacity is distributed across corridors and nodes. These approaches are conceptually related but analytically distinct. Connectivity does not necessarily imply dispersion, and resilience does not preclude structural dominance. The coexistence of small-world integration with heavy-tailed capacity allocation suggests that flow concentration can be interpreted as a distinct structural dimension within maritime systems.

Taken together, existing research suggests that maritime container systems are hierarchically organized, dynamically adaptive, and frequently concentrated in terms of capacity and traffic allocation. Topological indicators reveal integration patterns, resilience analyses highlight systemic sensitivities, and concentration metrics quantify dominance structures. These analytical traditions provide the conceptual foundation for examining how container flows are distributed across routes and nodes within bounded maritime systems.

\section{Analytical Framework}\label{sec:methodology}

This study represents the maritime container system as a set of FFE-weighted categorical distributions defined over routes, origin nodes, destination nodes, and industrial sectors. This representation treats container flows as distributions of volume across discrete categories, allowing structural properties such as concentration, dispersion, and asymmetry to be quantified in a consistent manner.

All metrics are computed within a unified probabilistic framework, in which total FFE volumes are normalized into share distributions that sum to one. This normalization enables the application of concentration and divergence measures defined on probability distributions.

Let $r$ index individual shipment records, and let $i \in \{1,\dots,n\}$ denote a category index (e.g., a route, port, or industry), where $n$ is the total number of categories. The total FFE mass associated with category $i$ is defined as
\begin{equation}
w_i = \sum_{r \in i} \mathrm{FFE}_r,
\label{eq:mass}
\end{equation}
where $w_i$ represents the total container volume assigned to category $i$. The corresponding share is defined as
\begin{equation}
s_i = \frac{w_i}{\sum_{j=1}^{n} w_j},
\qquad \sum_{i=1}^{n} s_i = 1,
\label{eq:share}
\end{equation}
where $s_i$ represents the proportion of total FFE associated with category $i$. The vector $\mathbf{s} = (s_1,\dots,s_n)$ therefore describes the distribution of container flows across categories, while $\mathbf{w} = (w_1,\dots,w_n)$ represents absolute volumes.

All subsequent measures are defined primarily as functions of the share vector $\mathbf{s}$, and, where required, of the corresponding mass vector $\mathbf{w}$ or its ordered form.

\subsection{Concentration Structure}

Research Questions~\ref{rq:route_concentration}, \ref{rq:port_dependency}, and \ref{rq:industry_specialization} require quantifying how container flows are distributed across categories. This subsection introduces complementary measures that capture different aspects of concentration and dispersion.

\subsubsection{Herfindahl--Hirschman Index}

The Herfindahl--Hirschman Index provides a summary measure of how strongly flows are concentrated in a small number of categories.

\begin{equation}
\mathrm{HHI} = \sum_{i=1}^{n} s_i^2.
\label{eq:hhi}
\end{equation}

The HHI \cite{herfindahl1997concentration,hirschman1980national} increases as larger shares are assigned to a smaller number of categories. Higher values indicate stronger concentration, meaning that a limited set of routes, ports, or industries accounts for most of the total container volume. Lower values indicate a more even distribution across categories.

\subsubsection{Concentration Ratios}

Concentration ratios provide a direct way to assess the contribution of the largest categories.

Let $s_{(1)} \ge s_{(2)} \ge \dots \ge s_{(n)}$ denote shares sorted in descending order. The $k$-concentration ratio is defined as

\begin{equation}
\mathrm{CR}_k = \sum_{i=1}^{k} s_{(i)}.
\label{eq:crk}
\end{equation}

This measure represents the cumulative share of the $k$ largest categories. For example, $\mathrm{CR}_3$ captures the share accounted for by the three largest routes or ports, making it useful for identifying dominance by a small subset of categories.

\subsubsection{Shannon Entropy}

While HHI emphasizes dominance, entropy captures the extent to which flows are spread across categories.

\begin{equation} 
H(\mathbf{s}) = -\sum_{i=1}^{n} s_i \ln s_i,
\label{eq:entropy}
\end{equation}

with normalized form \cite{shannon1948mathematical}

\begin{equation}
H_{\mathrm{norm}} = \frac{H(\mathbf{s})}{\ln(n)}.
\label{eq:entropy_norm}
\end{equation}

Entropy increases as the distribution becomes more even across categories. In this context, higher entropy indicates diversification, whereas lower entropy indicates concentration. This measure complements HHI by providing an alternative perspective on the same distribution.

\subsubsection{Gini Coefficient}

The Gini coefficient captures inequality in the distribution of volumes across categories.

Let $w_{(1)} \le \dots \le w_{(n)}$ denote ordered category totals, and let $C_i=\sum_{j=1}^{i} w_{(j)}$ denote cumulative volumes. The Gini coefficient is computed as \cite{gini1912variability,sen1997economic}

\begin{equation}
G = \frac{n+1 - 2\sum_{i=1}^{n}\left(\frac{C_i}{C_n}\right)}{n}.
\label{eq:gini}
\end{equation}

Higher values of $G$ indicate greater inequality, meaning that a small number of categories account for a large share of total volume, while many categories contribute only marginally.

\subsection{Directional Structural Asymmetry}

Research Question~\ref{rq:import_export_difference} examines whether import and export flows exhibit different distributional structures. This requires comparing two distributions defined over the same set of categories.

\subsubsection{Jensen--Shannon Distance}

The Jensen--Shannon distance provides a symmetric measure of divergence between two distributions.

Let $\mathbf{p}$ and $\mathbf{q}$ denote import and export share vectors. Define

\begin{equation}
\mathbf{m} = \frac{\mathbf{p}+\mathbf{q}}{2}.
\end{equation}

The Jensen--Shannon distance is defined as \cite{kullback1951information,lin2002divergence,jensen1906fonctions}

\begin{equation}
\mathrm{JSD}(\mathbf{p},\mathbf{q}) =
\sqrt{
\frac{1}{2} D_{\mathrm{KL}}(\mathbf{p}\|\mathbf{m}) +
\frac{1}{2} D_{\mathrm{KL}}(\mathbf{q}\|\mathbf{m})
}.
\label{eq:jsd}
\end{equation}

This measure takes values between zero and one. Values close to zero indicate similar distributions, while larger values indicate greater structural differences between import and export configurations.

\subsubsection{Rank Association}

In addition to distributional differences, it is useful to assess whether the relative ordering of categories is preserved.

Spearman's $\rho$ \cite{spearman1961proof} and Kendall's $\tau$ \cite{kendall1938new} are computed on aligned share vectors. These statistics measure the degree to which categories retain their relative ranking across imports and exports.

\subsection{Industry Directional Orientation}

To assess whether specific industries are more export-oriented or import-oriented, a relative measure is required.

\begin{equation}
R_i = \ln\left(
\frac{s_i^{\mathrm{export}} + \varepsilon}
{s_i^{\mathrm{import}} + \varepsilon}
\right),
\label{eq:logratio}
\end{equation}

where $s_i^{\mathrm{export}}$ and $s_i^{\mathrm{import}}$ denote the share of industry $i$ in exports and imports, respectively, and $\varepsilon>0$ prevents division by zero. Positive values indicate export orientation, while negative values indicate import orientation.

\subsection{Temporal Structural Stability}

Research Question~\ref{rq:temporal_change} examines how the structure of container flows evolves over time.

For each year $t$, distributions $\mathbf{s}_t$ are constructed. Structural change is measured relative to a base year $t_0$ as

\begin{equation}
\mathrm{Drift}_t = \mathrm{JSD}(\mathbf{s}_{t_0}, \mathbf{s}_t).
\label{eq:drift}
\end{equation}

This measure captures changes in the distribution of flows across categories independently of total volume changes. Spearman correlations between adjacent years are also used to assess the persistence of category rankings over time.

\section{Data Collection and Structure} \label{data_collection}

This study is based on shipment-level customs records covering the period 2019--2022. The data were obtained from the Mauritanian Customs Administration in spreadsheet format and consist of two direction-specific datasets (imports and exports). Each observation corresponds to a declared containerized shipment measured in FFE, which serves as the quantitative basis for all subsequent analysis.

To ensure analytical consistency, the raw datasets were harmonized into a unified structure. Column names were standardized across import and export files to allow consistent interpretation of variables. Only records containing all required analytical fields were retained.

Textual variables were normalized by trimming whitespace and converting values to uppercase. Empty and null-like entries were treated as missing values. The variables \textit{Year} and \textit{FFE} were converted to numeric format. Records with missing required fields or invalid numerical values were removed. Zero-volume observations were excluded, and no imputation or distributional smoothing procedures were applied.

Full-row duplicate observations were not removed, as no unique shipment identifier is available to distinguish true duplicates from valid repeated records. All observations are retained to preserve the structure of the original dataset.

For analytical consistency, origin and destination nodes were defined symmetrically across trade directions. For imports, the port of loading and port of discharge were mapped to \textit{Origin\_Node} and \textit{Destination\_Node}, respectively. For exports, the export loading port and place of delivery were mapped in an analogous manner. The cleaned datasets were then concatenated into a unified shipment-level dataset.

Each shipment is associated with a maritime service corridor classification recorded in customs documentation. These route codes correspond to predefined liner shipping corridor categories, summarized in Table~\ref{tab:routes}.

\begin{table}[h]
\centering
\caption{Maritime service route classification}
\label{tab:routes}
\begin{tabular}{p{2cm} p{11cm}}
\hline
Route & Service Description \\
\hline
W1 & Europe -- West Africa \\
W2 & Middle East -- West Africa \\
W3 & Asia -- West Africa (via Suez Canal) \\
W4 & Americas -- West Africa \\
W5 & South/East Africa -- West Africa \\
X6 & Intra-Africa / Short Sea \\
\hline
\end{tabular}
\end{table}

\begin{figure}[h]
\centering
\includegraphics[width=\textwidth]{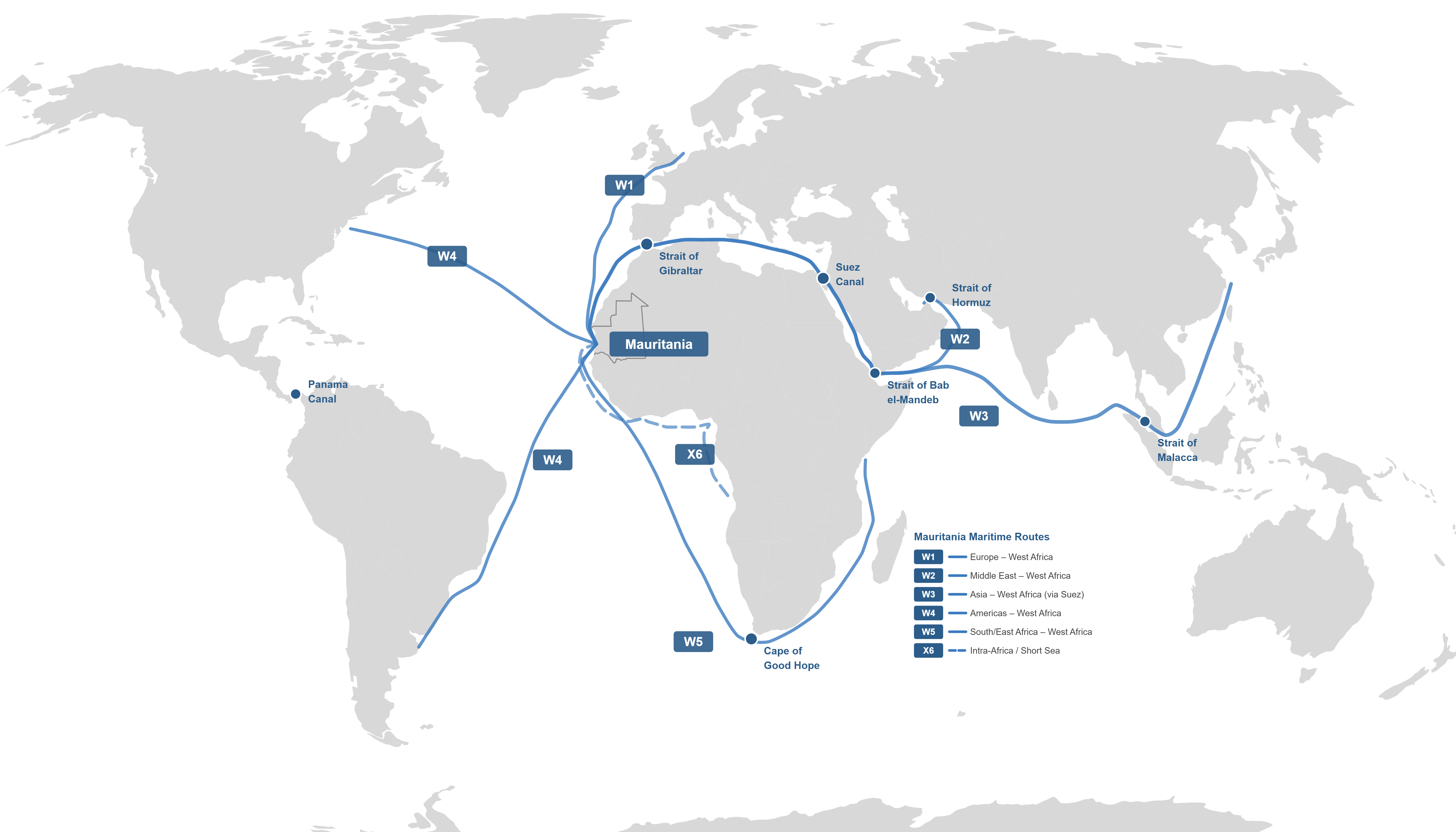}
\caption{Schematic representation of the main maritime route classes.}
\label{fig:route_scheme}
\end{figure}

Figure~\ref{fig:route_scheme} provides a schematic representation of these route categories. The figure illustrates the geographic orientation of the route taxonomy and does not represent observed flows.

\begin{figure}[h]
\centering
\includegraphics[width=\textwidth]{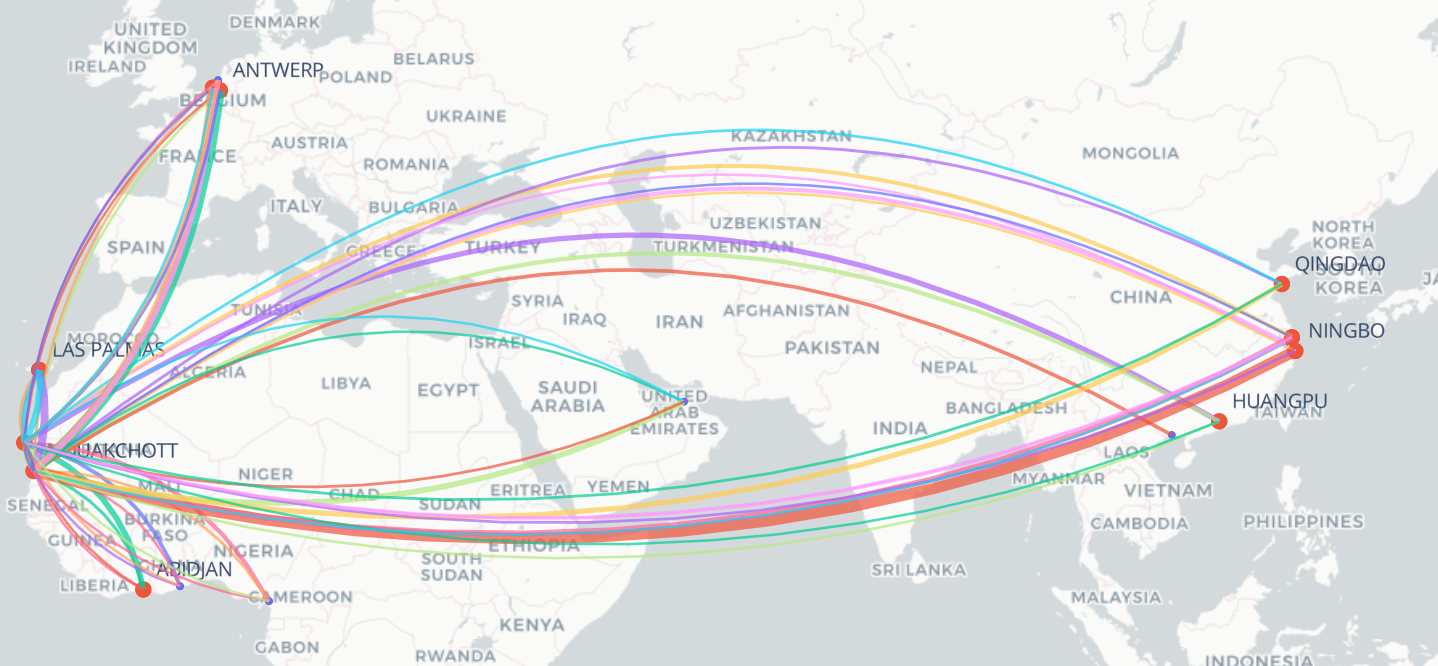}
\caption{Most active import and export corridors (2019--2022).}
\label{fig:mr_routes}
\end{figure}

Furthermore, Figure~\ref{fig:mr_routes} presents the most active observed corridors in the dataset, complementing the route classification with an empirical visualization. 

It is important to highlights that a small number of observations are labeled as ``UNKNOWN'' in the route field (six records totaling 5.5 FFE). In addition, two import observations (totaling 1.5 FFE) report Tangier Med as the port of discharge while Mauritania is recorded as the delivery country. These cases represent a negligible share of total volume and are retained without modification.

Table~\ref{tab:variables} summarizes the core variables used in the analysis.

\begin{table}[h]
\centering
\caption{Core variables and definitions}
\label{tab:variables}
\begin{tabular}{p{3cm} p{10cm}}
\hline
Variable & Definition \\
\hline
Year & Calendar year (2019--2022) \\
Direction & IMPORT or EXPORT \\
Route & Maritime corridor classification \\
Origin\_Node & Port of loading \\
Destination\_Node & Port of discharge or delivery \\
FFE & Shipment volume (weight variable) \\
Industry & Sector classification \\
Commodity & Declared commodity \\
\hline
\end{tabular}
\end{table}

All data preparation and computations were conducted in Python using \texttt{pandas}, \texttt{NumPy}, \texttt{SciPy}, and visualization using matplotlib and seaborn. 

\section{Exploratory Data Analysis} \label{eda}

The  dataset contains 105{,}686 shipment-level observations covering the period 2019--2022. Each record represents a containerized movement measured in FFE. The variables describe shipment timing, trade direction, maritime route classification, origin and destination nodes, industrial category, and shipment volume.

Directional composition is uneven, where 92{,}146 observations correspond to imports and 13{,}540 to exports. These counts represent shipment frequency and do not reflect volumetric shares.

\subsection{Distributional Properties of Shipment Volumes}

Table~\ref{tab:ffe_summary} reports descriptive statistics of shipment-level FFE. The median shipment size equals 1 FFE, while the maximum reaches 184 FFE. The mean (2.24) exceeds the median, and the standard deviation (4.59) reflects dispersion in shipment sizes. Most observations are concentrated at low values, with fewer high-volume shipments.

\begin{table}[H]
\centering
\caption{Shipment-Level FFE Distribution}
\label{tab:ffe_summary}
\begin{tabular}{lr}
\hline
Statistic & Value \\
\hline
Count & 105{,}686 \\
Mean & 2.23997 \\
Std. Dev. & 4.59360 \\
Minimum & 0.0868 \\
Median (50\%) & 1.0000 \\
75th Percentile & 2.0000 \\
Maximum & 184.0000 \\
\hline
\end{tabular}
\end{table}

\begin{figure}[H]
\centering
\includegraphics[width=0.8\textwidth]{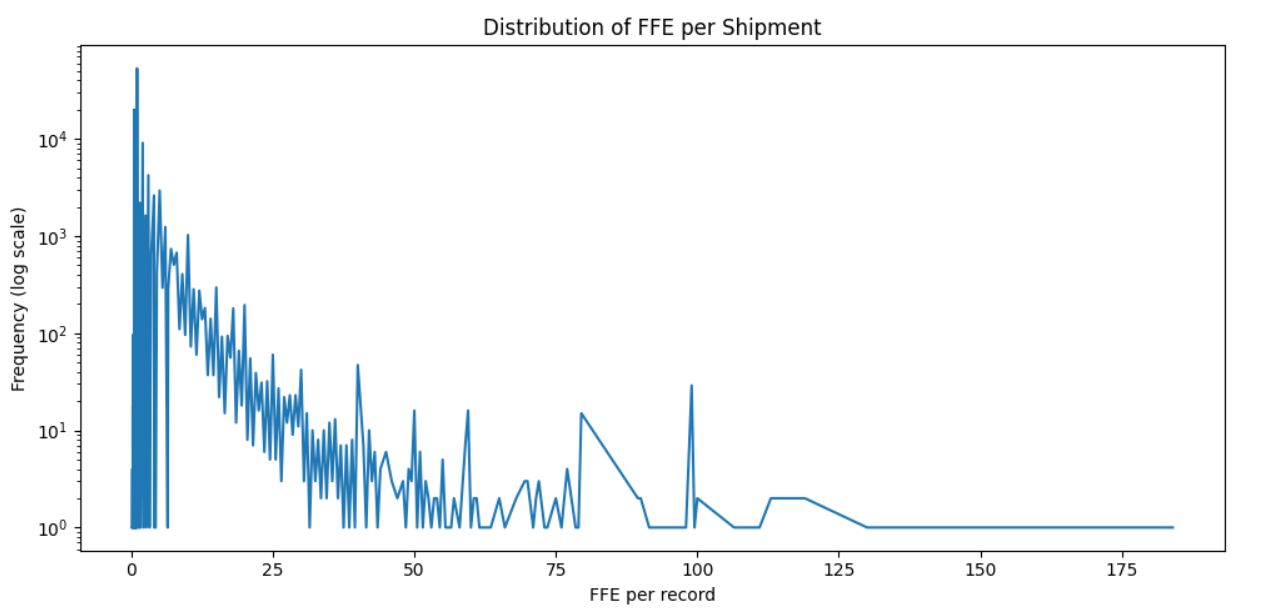}
\caption{Distribution of FFE per Shipment (Log Frequency Scale)}
\label{fig:ffe_distribution}
\end{figure}

Figure~\ref{fig:ffe_distribution} presents the distribution of shipment sizes on a logarithmic frequency scale. The distribution is highly skewed toward low FFE values, with a large concentration of observations at or near one container. As FFE increases, the frequency declines rapidly across several orders of magnitude, indicating that high-volume shipments occur infrequently relative to the bulk of observations. The visible upper tail extends to values above 100 FFE, confirming the presence of a limited number of high-volume shipments within the dataset.

\subsection{Temporal Flow Dynamics}

Table~\ref{tab:annual_totals} reports annual total FFE by direction. Import volumes increase from 41{,}678.5 FFE in 2019 to 46{,}778.5 FFE in 2021, followed by a decrease to 45{,}215.5 FFE in 2022. Export volumes decrease from 16{,}741.0 FFE in 2019 to 12{,}919.5 FFE in 2022. The magnitude of import flows exceeds export flows in all years, with imports consistently representing the dominant share of total volume.

\begin{table}[H]
\centering
\caption{Annual Total FFE by Direction}
\label{tab:annual_totals}
\begin{tabular}{lrr}
\hline
Year & Export & Import \\
\hline
2019 & 16{,}741.0 & 41{,}678.5 \\
2020 & 14{,}579.5 & 44{,}700.0 \\
2021 & 14{,}121.0 & 46{,}778.5 \\
2022 & 12{,}919.5 & 45{,}215.5 \\
\hline
\end{tabular}
\end{table}

\begin{figure}[H]
\centering
\includegraphics[width=0.8\textwidth]{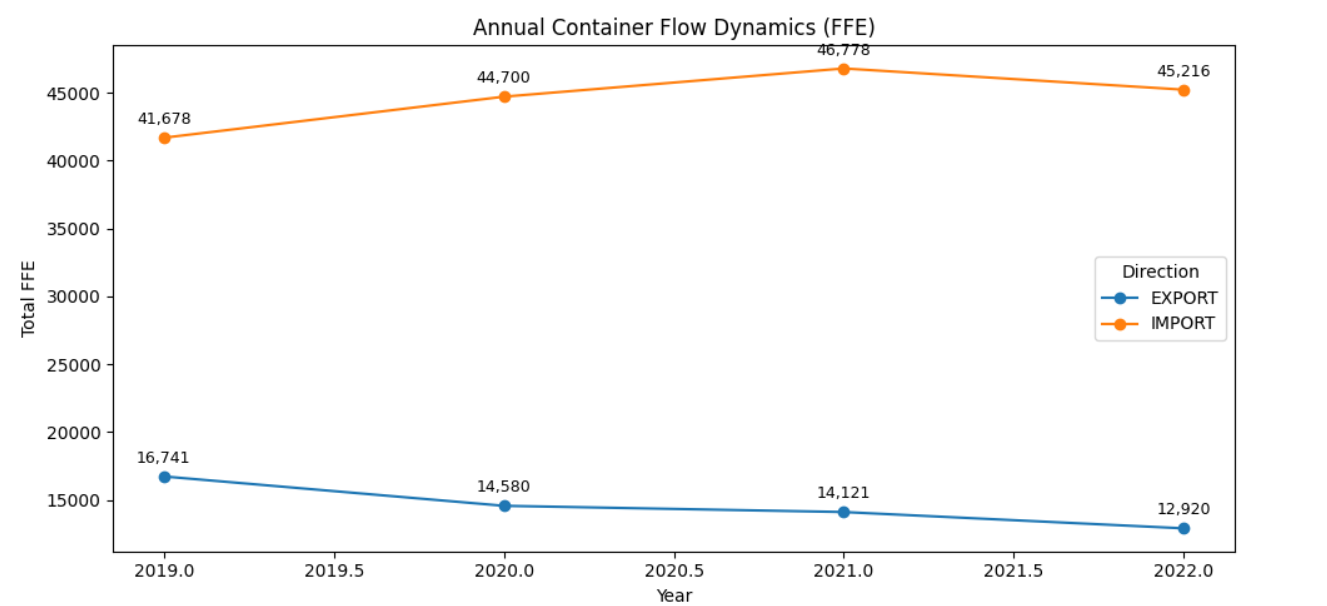}
\caption{Annual Container Flow Dynamics (FFE)}
\label{fig:annual_trend}
\end{figure}

Figure~\ref{fig:annual_trend} shows these patterns. Import flows follow a rising trend between 2019 and 2021 and then decline slightly, while export flows decrease more gradually over the entire period. The gap between import and export volumes remains substantial and persistent across all years.

\subsection{Route Concentration Patterns}

Table~\ref{tab:top_routes} reports total FFE and shares by maritime route. The two largest routes (W3 and W1) account for 38.16\% and 35.77\% of total FFE, respectively. Combined, they represent approximately 73.9\% of total volume. The third-largest route (W5) contributes 10.24\%, increasing the cumulative share of the three leading routes to approximately 84.2\%. The remaining routes individually account for less than 10\% of total FFE.

\begin{table}[H]
\centering
\caption{Top Maritime Routes}
\label{tab:top_routes}
\begin{tabular}{lrr}
\hline
Route & Total FFE & Share \\
\hline
W3 & 90{,}349.0 & 0.3816 \\
W1 & 84{,}678.0 & 0.3577 \\
W5 & 24{,}248.0 & 0.1024 \\
W2 & 22{,}576.0 & 0.0954 \\
W4 & 10{,}776.5 & 0.0455 \\
X6 & 4{,}100.5 & 0.0173 \\
UNKNOWN & 5.5 & 0.0000 \\
\hline
\end{tabular}
\end{table}

As explained in previous section, the category labeled ``UNKNOWN'' corresponds to six observations totaling 5.5 FFE and represents a negligible share of total volume.

\subsection{Industrial Composition}

Table~\ref{tab:top_industries} summarizes the distribution of flows across industrial categories. The largest category (Agriculture and Forestry) accounts for 17.1\% of total FFE, followed by ``Industry not classified'' (14.7\%) and frozen fish and seafood (13.2\%). The remaining categories contribute smaller shares, each below 11\% of total volume.

\begin{table}[H]
\centering
\caption{Top Industrial Categories}
\label{tab:top_industries}
\begin{tabular}{lrr}
\hline
Industry & Total FFE & Share \\
\hline
Agriculture \& Forestry & 40{,}502.5 & 0.1711 \\
Industry not classified & 34{,}742.0 & 0.1468 \\
Frozen Fish \& Seafood & 31{,}228.0 & 0.1319 \\
Food \& Beverage & 25{,}256.5 & 0.1067 \\
Finished Manufacturing & 15{,}739.5 & 0.0665 \\
\hline
\end{tabular}
\end{table}

The descriptive results show that shipment sizes are concentrated at low FFE values, total flows are dominated by imports, route distributions are concentrated in a small number of corridors, and industrial activity is distributed across several categories with varying shares. 

\section{Results and  Discussion}\label{sec:discussion}

This section answers the research questions by characterizing concentration, dependency, specialization, directional asymmetry, and temporal stability in containerized maritime system described by the dataset. The focus here is on structural metrics (concentration indices, divergence measures, and stability diagnostics) and on their interpretation in relation to network dependency and diversification.

\subsection{RQ1: Route Concentration}\label{sec:rq1}

Route concentration is evaluated using complementary concentration and inequality measures, including the HHI, concentration ratios (CR3, CR5), the Gini coefficient, and normalized entropy. Container flows are distributed across seven route categories, but the allocation of volume is uneven.

\begin{table}[H]
\centering
\caption{Route concentration metrics (overall, imports, exports).}
\label{tab:rq1_route_concentration}
\begin{tabular}{lrrrrrr}
\toprule
Scope & $n$ routes & Total FFE & HHI & CR3 & CR5 & Gini \\
\midrule
All & 7 & 236{,}733.5 & 0.2956 & 0.8418 & 0.9827 & 0.5379 \\
Imports & 7 & 178{,}372.5 & 0.3255 & 0.8999 & 0.9807 & 0.5861 \\
Exports & 6 & 58{,}361.0 & 0.3445 & 0.9840 & 0.9998 & 0.5540 \\
\bottomrule
\end{tabular}
\end{table}

As shown in Table~\ref{tab:rq1_route_concentration} for the combined system, the HHI equals 0.296, indicating a moderate-to-high level of concentration. The three largest routes account for 84.2\% of total FFE (CR3 = 0.842), while the five largest routes account for more than 98\% (CR5 = 0.983). The Gini coefficient (0.538) and normalized entropy (0.722) jointly reflect a distribution in which most volume is allocated to a limited number of routes, while the remaining categories contribute marginal shares.

Directional results exhibit differences in the structure of concentration. For imports, concentration is higher than in the aggregate system (HHI = 0.325; CR3 = 0.900), indicating that a large share of import volume is allocated to a small subset of routes. For exports, the HHI (0.345) is of similar magnitude, but concentration ratios are substantially higher (CR3 = 0.984; CR5 $\approx$ 1), reflecting the smaller number of route categories and the allocation of nearly all export volume to the leading corridors.

The relationship between HHI and concentration ratios differs across directions. While HHI captures the overall distribution of shares, concentration ratios reflect the cumulative dominance of the largest routes. In the export case, the high CR values combined with a moderate HHI reflect both strong dominance of leading routes and the influence of a limited number of categories.

\begin{table}[H]
\centering
\caption{Route shares (all directions).}
\label{tab:rq1_route_shares}
\begin{tabular}{lrr}
\toprule
Route & FFE & Share \\
\midrule
W3 & 90{,}349.0 & 0.3816 \\
W1 & 84{,}678.0 & 0.3577 \\
W5 & 24{,}248.0 & 0.1024 \\
W2 & 22{,}576.0 & 0.0954 \\
W4 & 10{,}776.5 & 0.0455 \\
X6 & 4{,}100.5 & 0.0173 \\
UNKNOWN & 5.5 & 0.0000 \\
\bottomrule
\end{tabular}
\end{table}

From the Table~\ref{tab:rq1_route_shares} the two largest routes (W3 and W1) account for 38.2\% and 35.8\% of total FFE, respectively. Combined, they account for approximately 73.9\% of the total volume. The addition of the third-largest route (W5) increases the cumulative share of the three leading routes to 84.2\%, consistent with the reported CR3 value. The category labeled ``UNKNOWN'' corresponds to six observations totaling 5.5 FFE and represents a negligible share of total volume. Its inclusion does not affect the concentration measures.

\subsection{RQ2: Node Dependency (Origin vs Destination)}\label{sec:rq2}

Node dependency is examined separately for (i) origin nodes (ports of loading) and (ii) destination nodes (ports of discharge), because each dimension represents a different structural role in the network. A system may contain many categories while still displaying concentration if a limited subset of nodes accounts for a large share of total volume.

As reported in Table~\ref{tab:rq2_origin_metrics}, origin nodes are numerous ($n=609$), while concentration remains low (HHI = 0.0539; CR3 = 0.3248). This indicates that flows are distributed across a broad set of origin nodes. However, the Gini coefficient is very high (0.9207), which shows that this distribution is highly unequal: a limited group of origins contributes a substantial share of total volume, whereas many other origins account for only marginal shares. The normalized entropy value (0.6033) is consistent with this pattern.

\begin{table}[H]
\centering
\caption{Origin-node dependency metrics (overall, imports, exports).}
\label{tab:rq2_origin_metrics}
\begin{tabular}{lrrrrrr}
\toprule
Scope & $n$ origins & Total FFE & HHI & CR3 & CR5 & Gini \\
\midrule
All     & 609 & 236{,}733.5 & 0.0539 & 0.3248 & 0.4395 & 0.9207 \\
Imports & 609 & 178{,}372.5 & 0.0338 & 0.2548 & 0.3475 & 0.9013 \\
Exports & 2   & 58{,}361.0  & 0.5672 & 1.0000 & 1.0000 & 0.1833 \\
\bottomrule
\end{tabular}
\end{table}

The directional results in Table~\ref{tab:rq2_top_origins} reveal a clear structural contrast. Import flows are spread across a large number of origin nodes, which explains the low concentration values. By contrast, export flows originate from only two nodes, and this limited number of categories mechanically produces high concentration ratios. Overall, the origin-node distribution is characterized by a small number of high-volume ports followed by a long tail of lower-volume nodes.

\begin{table}[H]
\centering
\caption{Top origin nodes (all directions).}
\label{tab:rq2_top_origins}
\begin{tabular}{lrr}
\toprule
Origin node & FFE & Share \\
\midrule
NOUADHIBOU & 39{,}997.5 & 0.1690 \\
NOUAKCHOTT & 18{,}600.5 & 0.0786 \\
NINGBO     & 18{,}291.0 & 0.0773 \\
LAS PALMAS & 13{,}752.0 & 0.0581 \\
ANTWERP    & 13{,}414.5 & 0.0567 \\
JEBEL ALI  & 9{,}362.0  & 0.0395 \\
VLISSINGEN & 7{,}158.0  & 0.0302 \\
SHANGHAI   & 5{,}126.5  & 0.0217 \\
ROTTERDAM  & 4{,}624.5  & 0.0195 \\
QINGDAO    & 4{,}053.5  & 0.0171 \\
\bottomrule
\end{tabular}
\end{table}

Destination nodes display a different configuration. Although the number of categories remains relatively large ($n=282$), concentration is substantially higher (HHI = 0.4845; CR3 = 0.7844), indicating that a large share of total volume is directed toward a limited number of destination nodes.

\begin{table}[H]
\centering
\caption{Destination-node dependency metrics (overall, imports, exports).}
\label{tab:rq2_dest_metrics}
\begin{tabular}{lrrrrrr}
\toprule
Scope & $n$ destinations & Total FFE & HHI & CR3 & CR5 & Gini \\
\midrule
All     & 282 & 236{,}733.5 & 0.4845 & 0.7844 & 0.8373 & 0.9664 \\
Imports & 3   & 178{,}372.5 & 0.8477 & 1.0000 & 1.0000 & 0.6113 \\
Exports & 281 & 58{,}361.0  & 0.0533 & 0.3394 & 0.4382 & 0.8896 \\
\bottomrule
\end{tabular}
\end{table}

Two observations, totaling 1.5 FFE, record Tangier Med as the port of discharge while Mauritania is listed as the delivery country. Because these cases represent a negligible share of total volume, they are retained without modification. The directional breakdown further shows that imports are concentrated in a very small number of destination nodes, whereas exports are distributed across a much broader set of foreign destinations and therefore exhibit lower concentration.

\begin{table}[H]
\centering
\caption{Top destination nodes (all directions).}
\label{tab:rq2_top_dests}
\begin{tabular}{lrr}
\toprule
Destination node & FFE & Share \\
\midrule
NOUAKCHOTT & 163{,}564.5 & 0.6909 \\
NOUADHIBOU & 14{,}838.5  & 0.0627 \\
HUANGPU    & 7{,}279.5   & 0.0308 \\
ABIDJAN    & 6{,}479.5   & 0.0274 \\
LAS PALMAS & 6{,}050.0   & 0.0256 \\
DOUALA     & 3{,}325.5   & 0.0140 \\
TEMA       & 2{,}441.0   & 0.0103 \\
SHANGHAI   & 2{,}432.0   & 0.0103 \\
JIAXING    & 2{,}223.5   & 0.0094 \\
SHIDAO     & 1{,}702.0   & 0.0072 \\
\bottomrule
\end{tabular}
\end{table}

Taken together, the comparison between origin and destination structures points to a clear directional asymmetry: origin nodes are more dispersed in terms of categories, whereas destination nodes are more concentrated in terms of volume allocation.

\subsection{RQ3: Industry Specialization}\label{sec:rq3}

Industry specialization is assessed by applying concentration measures to the distribution of flows across industrial categories, with imports and exports evaluated separately. The aggregate results, reported in Table~\ref{tab:rq3_industry_metrics}, provide a first overview of the structural differences between inbound and outbound flows.

\begin{table}[H]
\centering
\caption{Industry concentration metrics (overall, imports, exports).}
\label{tab:rq3_industry_metrics}
\begin{tabular}{lrrrrrr}
\toprule
Scope & $n$ industries & Total FFE & HHI & CR3 & CR5 & Gini \\
\midrule
All     & 35 & 236{,}733.5 & 0.0960 & 0.4498 & 0.6229 & 0.6996 \\
Imports & 35 & 178{,}372.5 & 0.0965 & 0.4610 & 0.6155 & 0.7010 \\
Exports & 21 & 58{,}361.0  & 0.3520 & 0.8592 & 0.9657 & 0.8555 \\
\bottomrule
\end{tabular}
\end{table}

At the aggregate level, industry concentration remains low (HHI = 0.0960), indicating that total flows are distributed across a relatively broad set of categories. As shown in Table~\ref{tab:rq3_industry_metrics}, imports follow an almost identical pattern (HHI = 0.0965; CR3 = 0.4610), which confirms a diversified allocation of volume across industries.

Exports, however, exhibit a structurally different configuration. According to Table~\ref{tab:rq3_industry_metrics}, the export HHI increases to 0.3520, while the concentration ratios rise sharply (CR3 = 0.8592; CR5 = 0.9657). This combination of high HHI and high concentration ratios indicates that export activity is concentrated within a limited subset of industries rather than being broadly distributed. The Gini coefficient (0.8555) reinforces this interpretation by confirming a high degree of inequality across export categories.

Taken together, the contrast between imports and exports reflects differences in how volume is allocated across industry categories, rather than differences in the number of categories alone.

Additional insight is obtained by examining the directional composition of flows, as reported in Table~\ref{tab:rq3_directional_profile}.

\begin{table}[H]
\centering
\caption{Selected industry shares (imports vs exports).}
\label{tab:rq3_directional_profile}
\begin{tabular}{lrr}
\toprule
Industry & Import share & Export share \\
\midrule
Frozen Fish \& Seafood      & 0.0002 & 0.5345 \\
Agriculture \& Forestry     & 0.1593 & 0.2072 \\
Metal in Secondary Form     & 0.0183 & 0.1175 \\
Industry not classified     & 0.1634 & 0.0960 \\
Food \& Beverage            & 0.1384 & 0.0098 \\
\bottomrule
\end{tabular}
\end{table}

The directional profile in Table~\ref{tab:rq3_directional_profile} highlights a pronounced asymmetry between imports and exports. Frozen Fish \& Seafood alone accounts for more than half of export volume (53.45\%), while contributing only a negligible share of imports, indicating a strong specialization of export activity in this category. Metal in Secondary Form also shows a substantially higher share in exports relative to imports, further supporting the presence of export-side concentration.

In contrast, categories such as Food \& Beverage, machinery, and manufactured goods exhibit substantially higher shares in imports, indicating that inbound flows are more diversified in their industrial composition. This divergence reinforces the interpretation derived from the concentration metrics.

The category labeled ``Industry not classified'', as shown in Table~\ref{tab:rq3_directional_profile}, represents a non-negligible share in both directions. Because this category aggregates heterogeneous activities, it reflects a limitation in the source classification rather than a coherent industry grouping. While concentration measures remain valid for the observed distribution, the presence of this residual category introduces ambiguity in the interpretation of the detailed industry structure.

\subsection{RQ4: Import--Export Structural Asymmetry}\label{sec:rq4}

Structural differences between imports and exports are evaluated by comparing their distributions over routes, origin nodes, destination nodes, and industries using Jensen--Shannon distance (JSD) and rank correlations. JSD is bounded between 0 and 1 (base 2), where higher values indicate greater divergence between distributions. The results, summarized in Table~\ref{tab:rq4_asymmetry}, provide a comparative view of asymmetry across structural dimensions.

Route distributions exhibit a moderate level of divergence (JSD = 0.4848), as reported in Table~\ref{tab:rq4_asymmetry}. Rank correlations are positive but not statistically significant (Spearman $\rho = 0.5714$, $p = 0.1802$; Kendall $\tau = 0.5238$, $p = 0.1361$), indicating that while some similarity in ranking exists, it is not strong enough to establish a consistent ordering across directions. This suggests partial correspondence in route importance alongside differences in relative shares.

In contrast, node distributions display the highest levels of divergence. As shown in Table~\ref{tab:rq4_asymmetry}, origin-node distributions are nearly maximally different (JSD = 0.9964), and destination-node distributions exhibit a similarly extreme level of divergence (JSD = 0.9985). The associated rank correlations are close to zero and not statistically significant for both origin and destination nodes, indicating that node rankings are largely unrelated between imports and exports. This pattern reflects a structural separation in the sets of active nodes across directions.

Industry distributions also exhibit substantial divergence (JSD = 0.7511), although their structure differs from that of nodes. According to Table~\ref{tab:rq4_asymmetry}, industry shares show strong positive and statistically significant rank correlations (Spearman $\rho = 0.6610$, $p = 1.5\times 10^{-5}$; Kendall $\tau = 0.5186$, $p = 3.0\times 10^{-5}$). This indicates that, despite differences in relative shares, the ordering of industries remains broadly consistent across imports and exports.

\begin{table}[H]
\centering
\caption{Directional asymmetry metrics (imports vs exports).}
\label{tab:rq4_asymmetry}
\begin{tabular}{lrrrr}
\toprule
Category & JSD (base 2) & Spearman $\rho$ ($p$) & Kendall $\tau$ ($p$) \\
\midrule
Route            & 0.4848 & 0.5714 (0.1802) & 0.5238 (0.1361) \\
Origin node      & 0.9964 & 0.0565 (0.1639) & 0.0468 (0.1631) \\
Destination node & 0.9985 & $-0.0417$ (0.4857) & $-0.0348$ (0.4824) \\
Industry         & 0.7511 & 0.6610 ($1.5\times 10^{-5}$) & 0.5186 ($3.0\times 10^{-5}$) \\
\bottomrule
\end{tabular}
\end{table}

Taken together, the results in Table~\ref{tab:rq4_asymmetry} show that the magnitude of divergence varies systematically across structural dimensions. Node distributions exhibit near-maximal divergence, reflecting fundamentally different sets of active nodes across directions. Industry distributions combine substantial divergence with strong rank association, indicating preserved ordering despite unequal shares. Route distributions, by contrast, display lower divergence and only partial correspondence, suggesting a weaker form of structural similarity between imports and exports.

\subsection{RQ5: Temporal Structural Stability}\label{sec:rq5}

Temporal stability is assessed by tracking yearly concentration metrics and by measuring distributional drift relative to a base year (2019) using JSD. The results are evaluated across routes, origin nodes, destination nodes, and industries to identify differences in structural persistence over time.

Route concentration remains highly stable over time, as shown in Table~\ref{tab:rq5_routes}. Annual HHI values vary within a narrow range, from 0.3034 in 2019 to 0.2953 in 2022, indicating minimal changes in the distribution of flows across routes despite small fluctuations in total volume and the number of active routes.

\begin{table}[H]
\centering
\caption{Route concentration over time (all directions).}
\label{tab:rq5_routes}
\begin{tabular}{lccc}
\toprule
Year & HHI & $n$ routes & Total FFE \\
\midrule
2019 & 0.3034 & 7 & 58{,}419.5 \\
2020 & 0.2936 & 7 & 59{,}279.5 \\
2021 & 0.2948 & 7 & 60{,}899.5 \\
2022 & 0.2953 & 6 & 58{,}135.0 \\
\bottomrule
\end{tabular}
\end{table}

Consistent with the concentration results, the corresponding JSD values remain low (0.076--0.095), confirming that route distributions experience only limited temporal drift relative to the base year.

Origin-node distributions exhibit moderate temporal variation. As reported in Table~\ref{tab:rq5_origins}, HHI decreases from 0.0671 in 2019 to approximately 0.0502 in subsequent years, while the number of categories remains consistently high.

\begin{table}[H]
\centering
\caption{Origin-node concentration over time (all directions).}
\label{tab:rq5_origins}
\begin{tabular}{lccc}
\toprule
Year & HHI & $n$ origins & Total FFE \\
\midrule
2019 & 0.0671 & 385 & 58{,}419.5 \\
2020 & 0.0525 & 398 & 59{,}279.5 \\
2021 & 0.0502 & 377 & 60{,}899.5 \\
2022 & 0.0502 & 391 & 58{,}135.0 \\
\bottomrule
\end{tabular}
\end{table}

The decline in HHI indicates a shift toward a more dispersed allocation across origin nodes. At the same time, distributional drift increases over time, with JSD rising from 0 in 2019 to 0.250 in 2022. This increase is driven primarily by import flows (up to 0.283), whereas export origin distributions remain highly stable (JSD $\approx$ 0.02--0.03).

Destination-node distributions combine increasing concentration with measurable temporal drift. As shown in Table~\ref{tab:rq5_dests}, HHI rises from 0.4321 in 2019 to approximately 0.5066 in 2022, while the number of categories declines from 183 to 149.

\begin{table}[H]
\centering
\caption{Destination-node concentration over time (all directions).}
\label{tab:rq5_dests}
\begin{tabular}{lccc}
\toprule
Year & HHI & $n$ destinations & Total FFE \\
\midrule
2019 & 0.4321 & 183 & 58{,}419.5 \\
2020 & 0.4934 & 178 & 59{,}279.5 \\
2021 & 0.5086 & 174 & 60{,}899.5 \\
2022 & 0.5066 & 149 & 58{,}135.0 \\
\bottomrule
\end{tabular}
\end{table}

JSD values increase to 0.207 by 2022, indicating growing divergence from the base-year distribution. This pattern differs by direction: import destinations remain highly stable (JSD below 0.02), whereas export destinations exhibit substantially higher drift, reaching 0.395.

Industry distributions display intermediate stability relative to routes and nodes. As reported in Table~\ref{tab:rq5_industry}, HHI remains within a narrow range (0.0932--0.1053), while the number of categories declines from 32 to 23 over the period.

\begin{table}[H]
\centering
\caption{Industry concentration over time (all directions).}
\label{tab:rq5_industry}
\begin{tabular}{lccc}
\toprule
Year & HHI & $n$ industries & Total FFE \\
\midrule
2019 & 0.1032 & 32 & 58{,}419.5 \\
2020 & 0.0940 & 31 & 59{,}279.5 \\
2021 & 0.0932 & 30 & 60{,}899.5 \\
2022 & 0.1053 & 23 & 58{,}135.0 \\
\bottomrule
\end{tabular}
\end{table}

Despite the stability in concentration, JSD values remain within a moderate range (0.216--0.228), indicating persistent variation in the relative shares of industries. This variation is more pronounced for imports (up to 0.251) than for exports (approximately 0.15--0.17).

Taken together, the results across Tables~\ref{tab:rq5_routes}--\ref{tab:rq5_industry} indicate a clear hierarchy of temporal stability. Route distributions are the most stable, node distributions exhibit greater variation over time, and industry distributions occupy an intermediate position between these two extremes.

\subsection{Synthesis}

The results provide a consistent structural characterization of Mauritania’s containerized maritime system across routes, nodes, industries, and time, based on concentration and divergence measures.

With respect to Research Question~\ref{rq:route_concentration}, route distributions are concentrated, with HHI equal to 0.296 and the three largest corridors accounting for approximately 84.2\% of total FFE. This indicates that flow allocation is organized around a limited number of dominant routes, which form the primary structural backbone of the system.

For Research Question~\ref{rq:port_dependency}, node structures differ markedly between origins and destinations. Origin nodes are numerous ($n=609$) and weakly concentrated (HHI = 0.054), while destination nodes exhibit substantially higher concentration (HHI = 0.485), with the majority of flows allocated to a small number of nodes. Directional decomposition shows an inversion: imports combine dispersed origins with highly concentrated destinations, whereas exports combine concentrated origins ($n=2$, CR3 = 1) with dispersed destinations ($n=281$).

Regarding Research Question~\ref{rq:industry_specialization}, industry distributions exhibit clear directional differences. Import flows are broadly distributed (HHI = 0.096), while export flows are strongly concentrated (HHI = 0.352), with the three largest industries accounting for approximately 85.9\% of export volume. This concentration is driven by a small set of dominant categories, including frozen fish and seafood.

For Research Question~\ref{rq:import_export_difference}, import and export distributions diverge across all structural dimensions. Divergence is highest for node distributions (JSD $\approx$ 0.996 for origins and 0.999 for destinations), indicating near-complete separation of node structures. Industry distributions also show substantial divergence (JSD = 0.751), while route distributions exhibit lower divergence (JSD = 0.485) and partial rank correspondence.

Finally, with respect to Research Question~\ref{rq:temporal_change}, temporal patterns differ across dimensions. Route distributions remain stable over time (HHI $\approx$ 0.29--0.30; JSD $< 0.10$), while node distributions exhibit higher variation, particularly for origin nodes (JSD up to 0.250) and export destinations (JSD up to 0.395). Industry distributions show intermediate behavior, with relatively stable concentration and moderate drift (JSD $\approx$ 0.21--0.23).

Taken together, the results show that the system is organized around stable corridor-level concentration, combined with direction-specific node configurations and persistent structural asymmetry between imports and exports.

\section{Conclusion}\label{sec:conclusion}

This study provides a distribution-based characterization of Mauritania’s maritime container system over 2019--2022 using shipment-level FFE data. By representing routes, nodes, and industries as categorical supports of FFE-weighted distributions, the analysis quantifies concentration, dependency, directional asymmetry, and temporal stability within a unified framework.

The results show that container flows are concentrated across a limited number of maritime corridors, with an overall HHI of 0.296 and the three largest routes accounting for approximately 84\% of total FFE. This concentration defines the primary allocation structure of the system and remains stable over time.

Node structures differ systematically across directions. Origin nodes are numerous and weakly concentrated (HHI = 0.054), while destination nodes exhibit substantially higher concentration (HHI = 0.485), with the majority of flows allocated to a small number of nodes. Directional patterns show an inversion: imports combine dispersed origins with highly concentrated destinations, whereas exports combine concentrated origins with dispersed destinations.

Industry distributions also differ between imports and exports. Import flows are broadly distributed (HHI = 0.096), while export flows are strongly concentrated (HHI = 0.352), with more than half of export volume concentrated in a single category (frozen fish and seafood).

Comparisons between import and export distributions confirm structural asymmetry across all dimensions. Divergence is most pronounced for node distributions (JSD $\approx$ 1), followed by industries (JSD = 0.751), while route distributions exhibit lower divergence (JSD = 0.485). These results indicate that asymmetry arises primarily from differences in node composition rather than from complete separation of corridor structures.

Temporal analysis indicates that concentration patterns remain stable at the corridor level (HHI $\approx$ 0.29--0.30; JSD $< 0.10$), while node distributions particularly export destinations (JSD up to 0.395) and import origins exhibit measurable variation over time. This suggests that structural change occurs primarily through reconfiguration of node-level connections.

The analysis is subject to several limitations. First, the results are distributional and do not identify causal mechanisms or operational performance characteristics. Second, the presence of an ``Industry not classified'' category reflects aggregation in sectoral coding, which may affect the measurement of specialization. Third, concentration measures depend on the categorical resolution of routes, nodes, and industries as defined in the dataset.

Future work may extend the analysis by refining category definitions, disaggregating residual industry classifications, and integrating external indicators of connectivity to contextualize the observed distributional patterns.


\section*{Data Availability}\label{sec:data-availability}
The shipment-level dataset used in this study contains proprietary customs information and is not publicly available. Researchers may inquire about data access by contacting the corresponding author or the Mauritanian Customs Administration, subject to applicable authorization and confidentiality requirements.

\section*{Declarations}\label{sec:declarations}

\subsection*{Funding}\label{sec:funding}
Not applicable.

\subsection*{Conflicts of Interest}\label{sec:conflicts}
The authors declare that there is no conflict of interest.

\subsection*{Consent for Publication}\label{sec:consent-publication}
All authors have reviewed and approved the final version of the manuscript and have provided consent for its publication.

\subsection*{Ethics Approval}\label{sec:ethics}
Not applicable.

\printbibliography

\end{document}